\documentclass[
 reprint,
 amsmath,
 amssymb,
 aps,
amssymb]{revtex4-2}
\usepackage{amssymb}
\usepackage{graphicx}
\usepackage{dcolumn}
\usepackage{bm}

\newcommand{\si}[1]{\textrm{#1}}

\newcommand{\kld}{\mathcal{D}}

\begin{document}

\title{Heating and Cooling are Fundamentally Asymmetric and Evolve Along
  Distinct Pathways} 
\author{Miguel Ib\'a\~nez$^{1,2}$, Cai Dieball$^3$, Antonio
  Lasanta$^{2,4,5}$, Alja\v{z} Godec$^{3,^\ast}$,\\\& Ra\'ul A. Rica$^{1,2,^\ast}$\\
\ \\
\normalsize{$^{1}$Universidad de Granada, Department of Applied Physics}\\
\normalsize{and Research Unit 'Modeling Nature' (MNat), Av. Fuentenueva s.n., 18071 Granada, Spain }\\
\normalsize{$^{2}$Nanoparticles Trapping Laboratory, Universidad de
  Granada, 18071 Granada, Spain}\\
\normalsize{$^{3}$Mathematical bioPhysics Group, Max Planck Institute for Multidisciplinary Sciences,}\\ \normalsize{37077 Göttingen, Germany}\\
\normalsize{$^{4}$Universidad de Granada, Department of Algebra,}\\
\normalsize{Facultad de Educaci\'on, Econom\'ia y Tecnolog\'ia de Ceuta, }\\
\normalsize{Universidad de Granada,
Cortadura del Valle, s/n, 51001 Ceuta, Spain}\\
\normalsize{$^{5}$Grupo de Teor\'ias de Campos y F\'isica Estad\'istica, Instituto Gregorio Mill\'an, }\\
\normalsize{Universidad Carlos III de Madrid}\\
\normalsize{Unidad Asociada al Instituto de Estructura de la Materia, CSIC, Spain}\\
\normalsize{$^\ast$To whom correspondence should be addressed;}\\
\normalsize{E-mail:   agodec@mpinat.mpg.de (AG); rul@ugr.es (RAR).}}


\date{\today}

\begin{abstract}
According to conventional wisdom, a system placed in
an environment with a different temperature tends to relax
to the temperature of the latter, mediated by the flows of heat and/or matter 
that are set solely by the temperature difference.
It is becoming clear, however, that
thermal relaxation 
is much
more intricate
when temperature
changes push the system far from thermodynamic equilibrium.
Interestingly, under such conditions heating was predicted to 
be faster than
cooling, which we experimentally confirm
using an optically trapped colloidal particle.
More strikingly, we show with both experiments and theory that between any pair of temperatures, heating is not only faster than cooling but the respective
processes in fact evolve along fundamentally distinct pathways,    
which we explain with a new theoretical framework we coin
``thermal kinematics''. 
\end{abstract}
\maketitle 
\textbf{Introduction}
The basic laws of thermodynamics dictate that any system in contact with an environment eventually relaxes to the temperature of its
surroundings as a result of irreversible flows that drive the system to
thermodynamic equilibrium. If the 
difference between the initial temperature of the system and that of the surroundings is small,
i.e. the system is initially ``close to equilibrium''
\cite{MazurGroot}, the relaxation is typically assumed to  
evolve  
quasi-statically through local 
equilibrium states --- 
an assumption that is  
justifiable only \emph{a posteriori} \cite{MazurGroot,Yokota}. 
However, if the temperature contrast is such that it pushes the system far from equilibrium, 
the assumption breaks down and the relaxation path is no longer unique, but
depends strongly on the initial condition. This gives rise to
counterintuitive phenomena, such as anomalous relaxation (also known as the
Mpemba effect) \cite{Lu2017MMpemba,Lasanta2017Mpemba,Baity2019MpembaSG,Kumar2020EMpemba,Carollo2021QMpemba,Kumar2022Anomalous,Klich} where, 
a system reaches equilibrium
faster upon a stronger temperature quench,
and the so-called Kovacs memory effect
\cite{Kovacs_granular,Kovacs_disordered,Kovacs_protein,Militaru2021Kovacs,Kovacs_liquid}
which features a non-monotonic evolution towards equilibrium. 



Intriguingly, thermal relaxation was recently predicted to depend also on the sign of the 
temperature change. Namely, considering two \emph{thermodynamically equidistant} (TE) temperatures ---
one higher and the other lower than an intermediate one selected such that the initial free energy difference with the equilibrium state 
is the same --- heating from the colder temperature was predicted to be faster
than cooling from the hotter one
\cite{Lapolla2020Faster,Van2021Faster}. This prediction challenges our
understanding of non-equilibrium thermodynamics as it compares
reciprocal relaxation processes elusive to classical thermodynamics. The initially hotter system must dissipate into the environment an excess of both, energy and entropy, whereas in the
colder system energy and entropy must increase \cite{Lapolla2020Faster,meibohm2021relaxation}. Moreover,
the comparison of heating and
cooling 
provokes an even more fundamental question, namely that of reciprocal
relaxation processes between two \emph{fixed} temperatures. According 
to the ``local equilibrium''
paradigm \cite{MazurGroot} 
the system relaxes quasi-statically and thus 
traces the same path along reciprocal processes. 
We show, however,
that this is \emph{not} the case: heating and cooling are inherently
asymmetric and evolve along distinct pathways. 

\begin{figure*}[hbtp]
	\centering
 	\includegraphics[width=\textwidth]{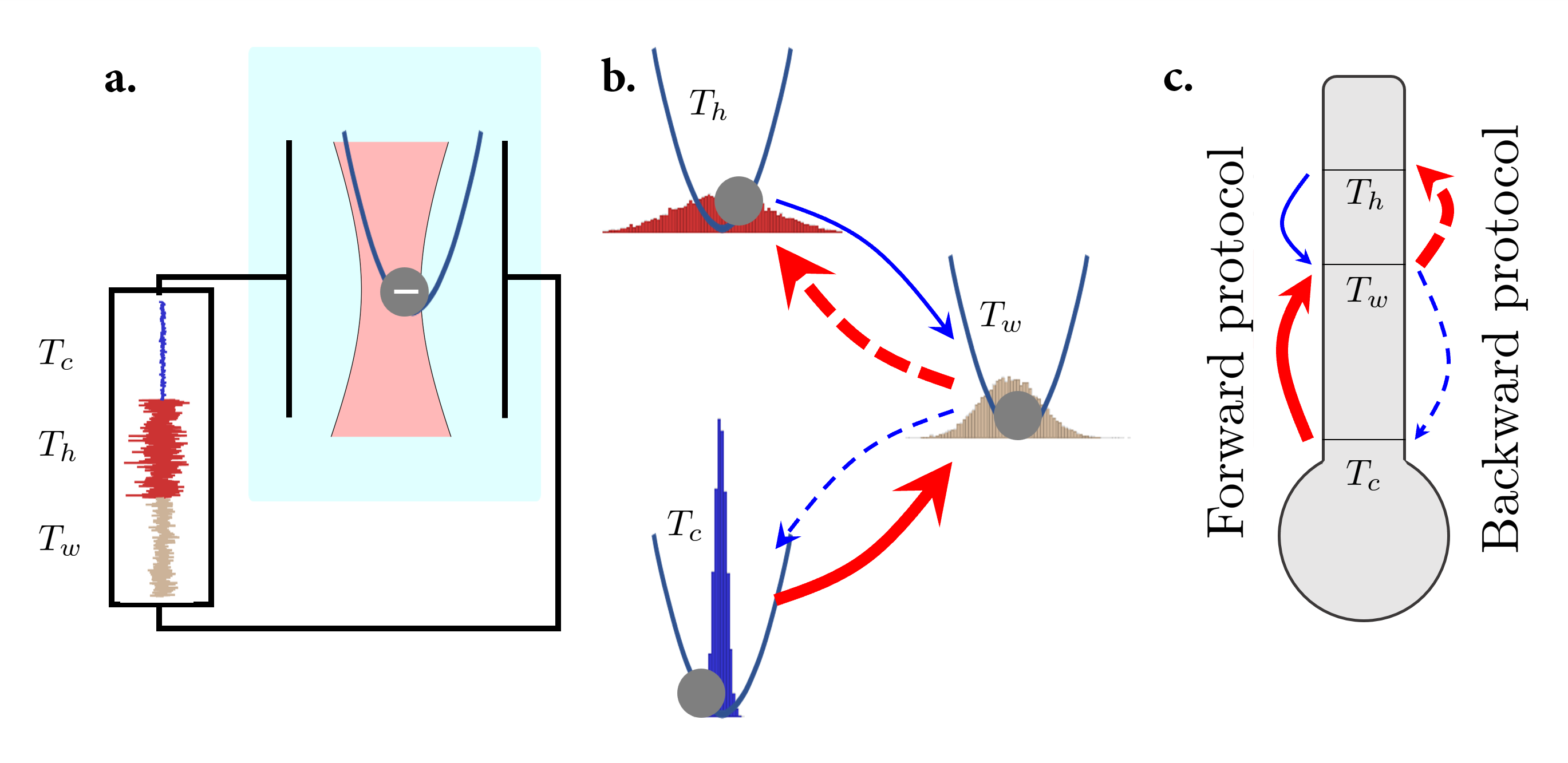}
    \caption{\textbf{Setup for probing the heating-cooling asymmetry.} \textbf{a.} Schematic representation of the experiment. A charged dielectric microparticle dispersed in water is confined in a parabolic trap generated by a tightly focused infrared laser. Its effective temperature is controlled by an electric field that shakes the particle, mimicking a thermal bath at a higher temperature than the water.  An arbitrary signal generator feeds a noisy signal with a Gaussian-white spectrum into a pair of gold microelectrodes immersed in the liquid, thus producing the required electric field. Therefore, the particle exhibits Brownian motion inside the trap, featuring a Gaussian distribution whose variance is determined by the effective temperature. \textbf{b.}  In experiments, we track the evolution of the position distribution upon quenches of the effective thermal bath during heating (red arrows) and cooling (blue arrows). \textbf{c.} Schematic representation of the respective protocols: in the forward protocol, the system is initially prepared at equilibrium with the thermal bath with a temperature higher ($T_h$) or lower ($T_c$) than the target ($T_w$) temperature. $T_h$ and $T_c$ are chosen to be thermodyncamically equidistant from $T_w$ with $T_h>T_w>T_c$. During the backward protocol the system relaxes at the respective thermodyncamically equidistant temperatures $T_h$ and $T_c$, starting from a common initial condition that is the equilibrium at $T_w$. In a third situation, only two temperatures are compared, considering the evolution of the system upon heating and cooling between them. In \textbf{b} and \textbf{c} 
    solid and dashed arrows stand for the forward and backward process, respectively, 
    and thick lines indicate faster evolution than thin ones.}
    \label{fig:fig1}
\end{figure*}

In this work, we use colloidal particles in temperature-modulated
optical traps to interrogate relaxation kinetics upon temperature
quenches (see Fig.~\ref{fig:fig1}), and unveil three fundamental asymmetries between
heating and cooling. We experimentally confirm the prediction that heating is faster than cooling in three complementary situations, precisely (i) that heating from a colder temperature towards an intermediate target temperature is faster
than cooling from the corresponding TE hotter temperature \cite{Lapolla2020Faster}. Unexpectedly, we also show (ii) that the reverse process, i.e. heating from the intermediate temperature to a hotter temperature is faster
than cooling 
to the corresponding TE colder temperature. Most
surprisingly, we show (iii) that between a fixed pair of temperatures, heating
is faster than the reciprocal cooling. In all cases, we provide mathematical proofs that establish these asymmetries as a general feature of systems with (at least locally) quadratic energy landscapes.\\
\indent 
A key result is that the production of entropy within the system
during heating is more efficient than heat dissipation during
cooling. Asymmetries (ii) and (iii) further imply that the
microscopic relaxation paths during heating and cooling are distinct. Moreover, whereas a system prepared at TE temperatures is by construction equally far from
equilibrium in terms of free energy, we show that the colder system is in fact statistically
farther from equilibrium and yet heating from said colder temperature is faster. Developing a new
framework we coin
``thermal kinematics'' we explain the asymmetry by means of the propagation in the space of probability distributions, which is intrinsically faster during heating.

\textbf{Heating and cooling at thermodynamically equidistant conditions}

Thermal relaxation kinetics beyond the ``local equilibrium'' regime
can be quantified within the framework of Stochastic
Thermodynamics \cite{Udo_2012,Sekimoto2010,Udo_2019} 
which requires the knowledge of statistics of all slow degrees of freedom. In the present work, where we use a colloidal particle with a diameter of 1 $\mu$m in a tightly focused laser (see Fig.~\ref{fig:fig1}), the overdamped regime ensures that only the position has to be analyzed \cite{Seifert2012Stochastic,blickle2012realization,martinez2015adiabatic}. Due to the symmetry of the tweezers setup, it suffices to follow a single coordinate of the particle as a function of time which we denote by $x_t$. We consider two different initial conditions. By the nature of the setup
(see Fig.~\ref{fig:fig1}) $x_t$ is initially in equilibrium in the
optical potential $U(x)$ at either the ``hot'', $T_h$, or ``cold'', $T_c$, temperature,
respectively, with a probability density $P^{j}_{\rm
  eq}(x)=\mathrm{e}^{(F_{j}-U(x))/k_{\rm B}T_j}$ where $F_{j}\equiv
-k_{\rm B}T_j\ln\int_{-\infty}^{\infty}\mathrm{e}^{-U(x)/k_{\rm
    B}T_{j}}dx$ is the equilibrium free energy at temperature $T_{j}$. 
 In the following, equilibrium probability densities are denoted by $P^{j}_{\rm eq}(x)$, where $j=h,w,c$ refers to the bath temperature $T_j$. Observables with both, subscript and superscript $i,f=h,w,c$, for example $A_i^f$, denote transient observables where the subscript refers to the initial and the superscript to the target state. 
The state of the system at any time is fully specified by, $P^{f}_i(x,t)$, the probability
density of the particle's position at time $t$.

\begin{figure*}[hbtp]
	\centering
 	\includegraphics{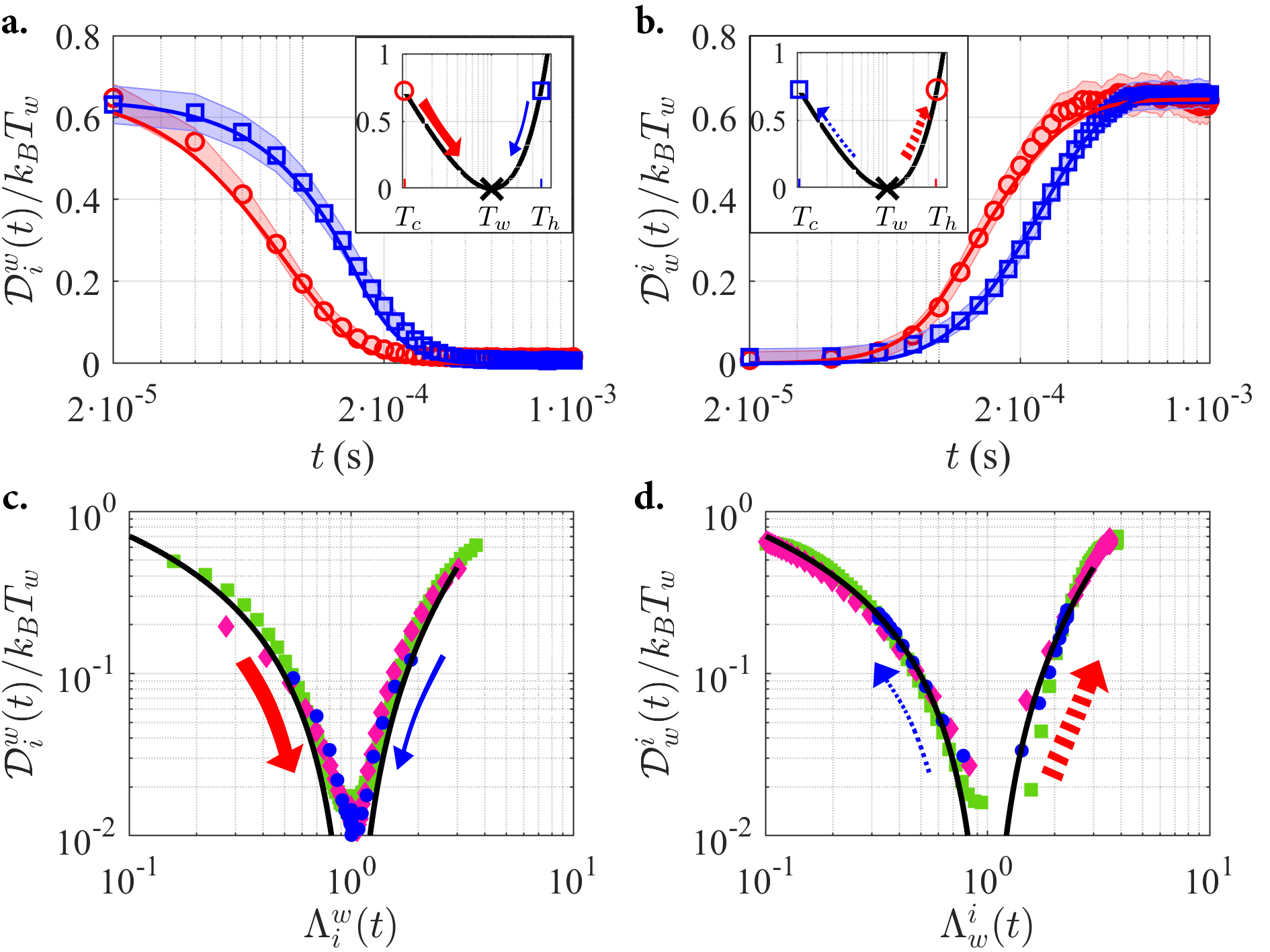}
    \caption{\textbf{Experimental evolution of the generalized excess free energy during heating and cooling at thermodynamically equidistant conditions.} Panels \textbf{a} and \textbf{c} correspond to the forward protocol and \textbf{b} and \textbf{d} to the backward counterpart. Thick, red arrows stand for heating while blue, thin arrows represent cooling. \textbf{a., b.} Time evolution of the generalized excess free energy for a characteristic time $\tau_c = \gamma/\kappa = 0.1844(3)$~ms, $T_c/T_w = 0.11(1)$, $T_h/T_w = 3.56(1)$. Red circles stand for heating and blue squares stand for cooling. Solid lines correspond to the theoretical predictions without fitting parameters. Insets represent the initial value of the relative entropy $\kld_i^w(0)/k_BT_w$ ($y$-axis) as a function of the temperature ($x$-axis) on the logarithmic scale. The arrows represent the evolution direction along the master curve $f(\rho)=(\rho-1-\ln \rho)/2$. The confidence regions have been estimated by quadratic uncertainty propagation from the standard deviation of the experimental histograms. \textbf{c., d.} Generalized excess free energy $\kld_{i/w}^{w/i}(t)/k_BT_w$ as a function of $\Lambda_{i/w}^{w/i}(t)$, along the master curve $f(\rho)$, for several different TE conditions. The corresponding time series are included in the Supplemental Material (Fig. S5).}
    \label{fig:fig2}
\end{figure*}

We first focus on the forward protocol where the relaxation occurs at the ``warm'' temperature $T_w$ and  $i=h,c$. The dynamics is ergodic and therefore
$P^w_{i}(x,t)$ relaxes towards $P^{w}_{\rm
  eq}(x)$.
We use $\langle \ldots\rangle^w_{i}$ to
denote averages over $P^{w}_i(x,t)$ and quantify the instantaneous
displacement from the equilibrium distribution $P^{w}_{\rm eq}(x)$ 
by means of the \emph{generalized excess free energy} given by
\cite{LEBOWITZ19571,Vaikuntanathan_2009,Lapolla2020Faster}
$\mathcal{D}^{w}_{i}(t)=\langle U(x)\rangle^w_{i}-k_{\rm
  B}T_{w}\langle \ln P^{w}_i(x,t)\rangle^w_{i}-F_{w}=k_{\rm
  B}T_{w}D[P^{w}_i(x,t)||P^w_{\rm eq}(x)]$
for $i=h,c$, where $D[P||Q]=\int P\ln(P/Q)dx$ is the relative entropy between the probability distributions $P$ and $Q$. Temperatures $T_h$
and $T_c$ are said to be TE from $T_w$
when the
initial excess free energies are equal, i.e.,
$\mathcal{D}^w_{h}(0)=\mathcal{D}^w_{c}(0)$
\cite{Lapolla2020Faster}.  
The unexpected prediction was made in Ref.~\cite{Lapolla2020Faster} that
$\mathcal{D}^w_{c}(t)<\mathcal{D}^w_{h}(t)$ at all times $t>0$. That is,
the system heats up to the temperature of its surroundings faster than
it cools down. 
Albeit an asymmetric relaxation is counter-intuitive, our experiments quantitatively corroborate this prediction to be true, see
Fig.~\ref{fig:fig2}. 

What may be even more surprising,
heating also turns out to be faster along the reversed,
\emph{backward} protocol. That is, we prepare the system to be in equilibrium at the
``warm'' temperature $T_w$
and track the relaxations at $T_h$ and $T_c$,
respectively. Likewise, we quantify the kinetics via the relative entropy
$\mathcal{D}^i_{w}(t)\equiv k_{\rm
  B}T_{w}D[P^i_{w}(x,t)||P^w_{\rm eq}(x)]$, such that
$\mathcal{D}^h_{w}(t)$ and 
$\mathcal{D}^c_{w}(t)$ evolve from zero and asymptotically converge to 
$\mathcal{D}^{h/c}_{w}(\infty)=\mathcal{D}^w_{h/c}(0)$. 
Although
$\mathcal{D}^i_{w}(t)$ sensibly quantifies the departure from
$P^w_{\rm eq}(x)$, it is strictly speaking \emph{not} an excess free energy,
in contrast to $\mathcal{D}^w_{i}(t)$ defined above, because $T_w$ and $P^w_{\rm eq}(x)$ no longer refer to the target equilibrium.  
We observe in Fig.~\ref{fig:fig2} that
$\mathcal{D}^h_{w}(t)>\mathcal{D}^c_{w}(t)$ 
for all times $t>0$, i.e.\ the system heats up to the new equilibrium at
$T_h$ faster than it cools back to $T_c$. This observation is
remarkable as it shows that heating is inherently faster than cooling
at TE conditions.\\

\indent To confirm these observations theoretically, we assume that the
particle's dynamics evolve in a parabolic potential with stiffness
$\kappa$, $U(x)=\kappa x^2/2$, according to the overdamped Langevin
equation $d x_t=-(\kappa/\gamma) x_t dt+d\xi^i_t$ with friction constant
$\gamma$ given by the Stokes' law $\gamma=6\pi r\eta$ where $\eta$ is
the viscosity of water. The thermal noise $d\xi^i_t$, where $i=h,w,c$
denotes the temperature of the reservoir,
vanishes on average 
and obeys the
Fluctuation-Dissipation Theorem 
$\langle d\xi^i_t
d\xi^i_{t'}\rangle=2(k_{\rm B}T_i/\gamma)\delta(t-t')dtdt'$.
Under these
assumptions we determine TE temperatures
$T_h$ and $T_c(T_h)$, i.e., we calculate $T_c$ after we arbitrarily set $T_h$, (see Eq.~(S6) in the SM)
and $\mathcal{D}^{w/i}_{i/w}(t)$ reads
\begin{equation}
\mathcal{D}^{w/i}_{i/w}(t) =
\frac{k_{\rm B}T_w}{2}[\Lambda^{w/i}_{i/w}(t)-1-\ln \Lambda^{w/i}_{i/w}(t)],
\label{KLD}
\end{equation}  
where
$\Lambda^{w}_{i}(t)=1+(T_{i}/T_w-1)\mathrm{e}^{-2(\kappa/\gamma)t}$ and
$\Lambda^{i}_{w}(t)=T_{i}/T_w+(1-T_{i}/T_w)\mathrm{e}^{-2(\kappa/\gamma)t}$ .
We consider $\mathcal{D}^{w}_{i}(t)$ during the forward and
$\mathcal{D}^{i}_{w}(t)$ during the backward protocol. 
According to Eq.~\eqref{KLD}, by plotting $\mathcal{D}^{w/i}_{i/w}(t)/k_{\rm
  B}T_w$ as a function of $\rho=\Lambda^{w/i}_{i/w}(t)$ all data should
collapse onto the master curve $f(\rho)=(\rho - 1-\ln\rho)/2$, which is
indeed what we observe in Fig.~\ref{fig:fig2}. Having
established the validity of the model, we prove (see Theorem 1 in SM) that our
observations hold for all
TE temperatures and for any
$\kappa$ and $\gamma$, i.e.\ 

\begin{equation}
\mathcal{D}^{w}_{c}(t)<\mathcal{D}^{w}_{h}(t)\quad{\rm
  and}\quad\mathcal{D}^{h}_{w}(t)>\mathcal{D}^{c}_{w}(t),\quad\text{for
  all}\,\, 0<t<\infty.
\label{asymmetry_1}
\end{equation}

Our observations in Fig.~\ref{fig:fig2} and the
inequalities \eqref{asymmetry_1} establish rigorously that at
TE conditions heating is
faster than cooling.\\
\indent Notwithstanding, these results are still unsatisfactory for two
reasons. First, $\mathcal{D}^{i}_{w}(t)$, unlike $\mathcal{D}_{i}^{w}(t)$, lacks a consistent
thermodynamic interpretation and, in particular, is not an excess free
energy. Second, neither the relative entropy $D[P||Q]$ nor $\sqrt{D[P||Q]}$ are a true metric; they are not symmetric, $D[P||Q]\ne D[Q||P]$, and do not
satisfy triangle inequalities. The latter in particular
implies that while a Pythagorean theorem holds (see
\cite{shiraishi2019information}; see SM for an experimental validation),
$D[P^i_{\rm eq}(x)||P^w_{\rm eq}(x)]\ge D[P^i_{\rm
    eq}(x)||P^i_w(x,t)]+D[P^i_w(x,t)||P^w_{\rm eq}(x)]$, where we used
$P^i_w(x,0)=P^i_{\rm eq}(x)$, the triangle inequality does
not, i.e., in general $\sqrt{D[P^i_{\rm eq}(x)||P^w_{\rm eq}(x)]}\not\le\sqrt{D[P^i_{\rm
    eq}(x)||P^i_w(x,t)]}+\sqrt{D[P^i_w(x,t)||P^w_{\rm eq}(x)]}$. 
As a result,
$\mathcal{D}^{w/i}_{i/w}(t)$ does not measure
``distance'' from equilibrium and
$\partial_t \mathcal{D}^{w/i}_{i/w}(t)$ is therefore not a ``velocity'', which
seems to preclude a kinematic description of relaxation. In the following sections, we show that combining Stochastic Thermodynamics with Information Geometry
\cite{Ito2020Stochastic} makes it possible to formulate a
\emph{thermal kinematics}.

\begin{figure*}[hbtp]
	\centering
	\includegraphics{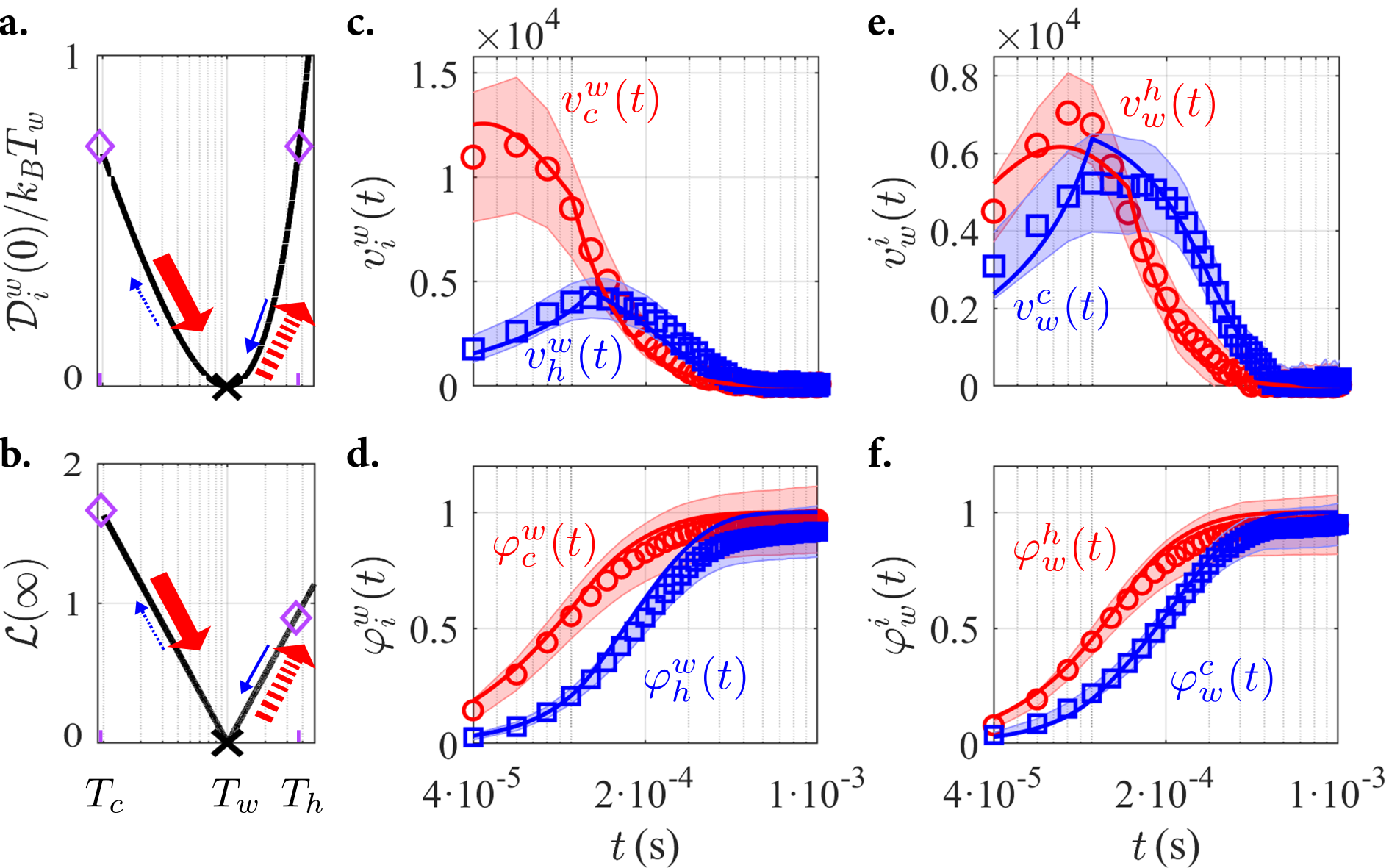}
    \caption{\textbf{Thermal kinematics of heating and cooling processes at thermodynamically equidistant conditions.} All data corresponds to the series shown in Fig.~\ref{fig:fig2}. As in previous figures, red arrows stand for heating while blue ones represent cooling, solid and dashed arrows refer to forward and backward protocol, respectively, and thicker lines indicate a faster evolution than thin ones. \textbf{a.} Initial value of the relative entropy $\kld_i^w(0)/k_BT_w$ as a function of the temperature. \textbf{b.} Total traversed statistical distance $\mathcal{L}_i^w(\infty)=\mathcal{L}_w^i(\infty)=\mathcal{L}(\infty)$ as a function of the temperature. \textbf{c-f} Temporal evolution of the instantaneous statistical velocity $v_i^w(t)$ (\textbf{c} and \textbf{e}) and the degree of completion $\varphi_i^w(t)$ (\textbf{d} and \textbf{f}) during the forward (\textbf{c} and \textbf{d}) and backward (\textbf{e} and \textbf{f}) protocol. Red circles stand for heating while blue squares correspond to cooling. Solid lines are theoretical predictions without fitting parameters. Confidence regions have been estimated by quadratic uncertainty propagation from the standard deviation of the experimental histograms.}
    \label{fig:fig3}
\end{figure*}

\textbf{Towards thermal kinematics}
Immense progress has been made in recent years in understanding non-equilibrium systems. The discovery of thermodynamic uncertainty relations
\cite{barato2015thermodynamic,gingrich2016dissipation,Dechant_PRX,dieball2022direct},
and ``speed limits''
\cite{okuyama2018quantum,shiraishi2018speed,shiraishi2019information,Ito2020Stochastic}
revealed that the entropy production rate, which quantifies irreversible
local flows in the system, universally bounds
fluctuations and the rate of change, respectively, in a
non-equilibrium system. Closely related is the so-called Fisher
Information 
known from Information Geometry, which quantifies how local flows
change in time and allows for defining a \emph{statistical distance}
\cite{Ito2020Stochastic,Crooks2007Measuring,Cover2006Elements}.\\ 
\indent In our context of thermal relaxation, an infinitesimal
statistical line element may be defined as follows. Since
$D[P_i^f(x,t+dt)||P_i^f(x,t)]=I_i^f(t)dt^2+\mathcal{O}(dt^3)$
(see SM, Eq.~(S16)), where
we introduced the Fisher
information $I_i^f(t)\equiv\langle(\partial_t\ln P_i^f(x,t))^2
\rangle_i^f$, we can define the line element as
$dl\equiv\sqrt{D[P_i^f(x,t+dt)||P_i^f(x,t)]}=\sqrt{I_i^f(t)}dt$ and
thus $v_i^f(t)\equiv \sqrt{I_i^f(t)}$ is the \emph{instantaneous
statistical velocity} of the system \cite{Ito2020Stochastic} relaxing from $P^i_{\rm eq}(x)$
at temperature $T_f$ towards $P^f_{\rm eq}(x)$. The statistical length traced by
$P_i^f(x,\tau)$ until time $t$ is
$\mathcal{L}^f_i(t)=\int_0^tv_i^f(\tau)d\tau$ and the distance between
initial and final states is thus given by $\mathcal{L}_i^f(\infty)$
and does \emph{not} depend
on the direction,
i.e.\ $\mathcal{L}_a^b(\infty)=\mathcal{L}_b^a(\infty)$, for two different temperatures $T_a$ and $T_b$. To establish
a kinematic basis for quantifying thermal relaxation kinetics we
define the \emph{degree of completion}
$\varphi^{w/i}_{i/w}(t)\equiv\mathcal{L}^{w/i}_{i/w}(t)/\mathcal{L}^{w}_{i}(\infty)$,
which increases monotonically between 0 and 1.\\
\indent Assuming that the system evolves according to overdamped Langevin
dynamics in a parabolic potential, we find (see Sec.~A.5 in SM)
$\mathcal{L}^{w}_{i}(\infty)=|\ln(T_i/T_w)|/\sqrt{2}$ and  
 \begin{eqnarray}
\varphi^{w/i}_{i/w}(t)&=&1-\frac{\ln(1+(T_{i/w}/T_{w/i}-1)\mathrm{e}^{-2(\kappa/\gamma)t})}{\ln(T_{i/w}/T_{w/i})}.
\label{DOFs}
\end{eqnarray}
Moreover, we prove (see Theorem 2 in SM) for any pair of
TE temperatures $T_h,T_c$ that
$\mathcal{L}^{w}_{c}(\infty)>\mathcal{L}^{w}_{h}(\infty)$ and yet
\begin{equation}
\varphi^{w}_{c}(t)>\varphi^{w}_{h}(t)
\quad\text{and}\quad\varphi_w^h(t)>\varphi_w^c(t)
\quad\text{for
  all}\,\,0<t<\infty.
\label{asymmetry_1a}
\end{equation}  
That is, the colder system is statistically farther from equilibrium
than the hotter system, but nevertheless,
heating is faster than cooling. On the one hand, Eq.~\eqref{asymmetry_1a}
confirms the asymmetry \eqref{asymmetry_1} from a kinematic point of
view. On the other hand, it reveals something more striking; during
heating, the system traces a longer path in the space of probability
distributions but it does so faster. The reason lies in the
propagation speed $v^{i/w}_{w/i}(t)$ at short times that is intrinsically
larger during heating than during cooling. This speed-up is due to
the entropy production in the system during heating being more
efficient than heat flow from the system to the environment during
cooling (see also \cite{Lapolla2020Faster}). These predictions are fully confirmed by experiments (see Fig.~\ref{fig:fig3}). The results show that an initial overshoot in $v^{w}_{c}(t)$ and $v^{h}_{w}(t)$ ensures that under TE conditions heating is, in both protocols, at all times faster than cooling according to the inequalities \eqref{asymmetry_1a}. Since both processes relax to the same equilibrium, $v^{w}_{c}(t)$ and $v^{h}_{w}(t)$ eventually must cross.

\begin{figure}
    \centering
	\includegraphics[width=0.8\linewidth]{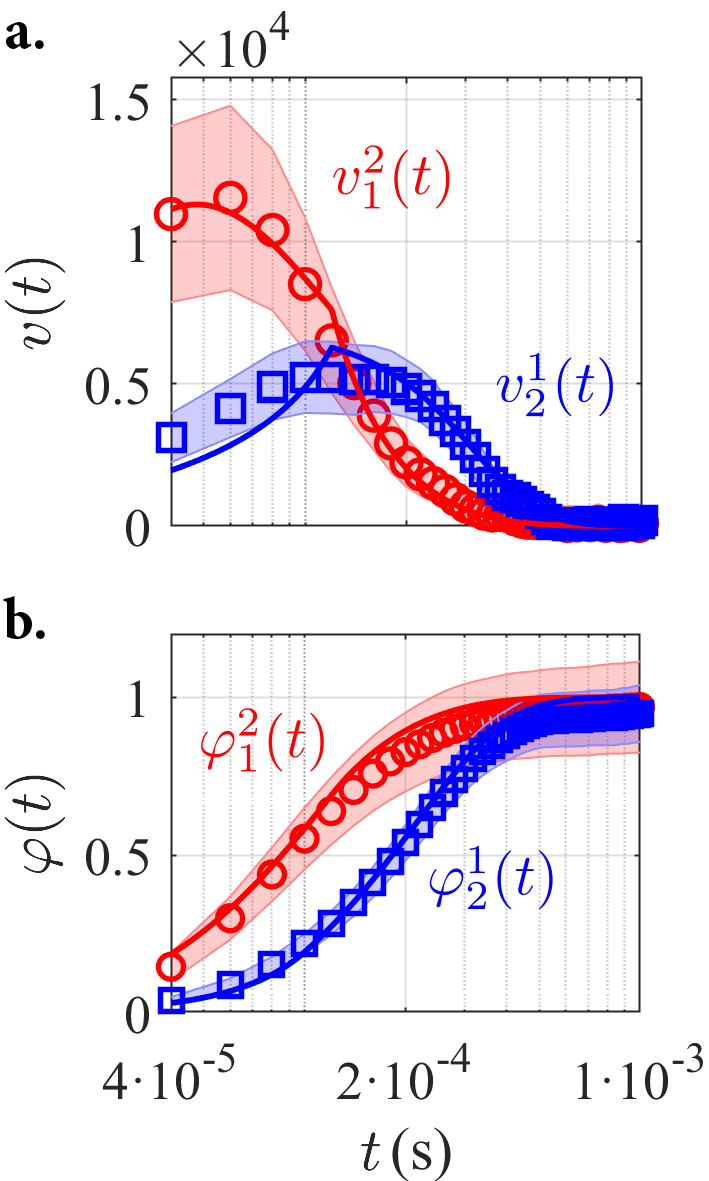} 
    \caption{\textbf{Thermal kinematics of heating and cooling between any pair of temperatures.} The data shown corresponds to $T_1=302(3) \ \si{K}$ and $T_2=2753(7) \ \si{K}$, and a characteristic time $\tau_c=\gamma/\kappa = 0.1844(3) \ \si{ms}$. \textbf{a.} Instantaneous statistical velocity $v(t)$ as a function of time. \textbf{b.} Degree of completion $\varphi(t)$ as a function of time. Red circles stand for heating, while blue squares correspond to cooling. Solid lines are theoretical predictions without fitting parameters. The confidence regions have been estimated by quadratic uncertainty propagation from the standard deviation of the experimental histograms.}
    \label{fig:fig4}
\end{figure}

\textbf{Heating between any pair of temperatures is faster than
  cooling} We now take our thermal kinematics approach one step further and consider two arbitrary fixed temperatures
$T_1<T_2$ and observe heating, i.e\ relaxation at $T_2$ in a
temperature quench from an
equilibrium prepared at $T_1$, and the reverse cooling, i.e.\ relaxation at $T_1$ in a
temperature quench from the
equilibrium at $T_2$. By construction, the distance between
initial and final states along the reciprocal processes is the \emph{same},
$\mathcal{L}^1_2(\infty)=\mathcal{L}^2_1(\infty)$. Nevertheless, according to our
model, in particular Eqs.~\eqref{DOFs}, we have for any $T_1<T_2$ (see Theorem 3 in SM) that
\begin{equation}
\varphi^2_1(t)>\varphi^1_2(t)\quad\text{for all}\quad 0<t<\infty.
\label{asymmetry_2}  
\end{equation}
That is, between any pair of temperatures heating is faster than cooling, which is a much stronger statement. Notice that it was not possible to make such a statement based on the generalized excess free energy, since in this case, the description of the backward process lacked physical consistency. This result highlights that
heating and cooling are inherently 
asymmetric processes and that this is neither limited to the
TE setting nor to strong quenches. The asymmetry
\eqref{asymmetry_2} is fully corroborated by experiments (see
Fig.~\ref{fig:fig4}). As before, it emerges due to an initial overshoot
in $v^{2}_{1}(t)$. However, in this case, the difference in velocities implies that the pathway taken during heating is fundamentally different from the pathway followed
during the reciprocal cooling process.

\textbf{Near equilibrium, heating and cooling become symmetric} Finally, we show that for quenches near equilibrium heating and cooling are indeed almost symmetric, in agreement with linear non-equilibrium thermodynamics
\cite{MazurGroot}. First, for $T_+=(1+\varepsilon)T_w$ with $0<\varepsilon\ll 1$, 
we find that
$T_-(T_+)=(1-\varepsilon+\mathcal{O}(\varepsilon^2))T_w$ (see Eq.~(S33) in SM), i.e., near
equilibrium, TE temperatures are
approximately equidistant
from the ambient temperature $T_w$. Second, we find in this limit
(see SM, Corollary 4) that 
$\varphi^i_w(t)=\varphi^w_i(t)=1-\mathrm{e}^{-2(\kappa/\gamma)t}+\mathcal{O}(\varepsilon)$ and
\begin{equation}
\varphi^c_w(t)=\varphi^h_w(t)+\mathcal{O}(\varepsilon)\quad\text{and}\quad\varphi^w_h(t)=\varphi^w_c(t)+\mathcal{O}(\varepsilon).
\end{equation}
That is, for near-equilibrium quenches heating and cooling are approximately
symmetric and the asymmetry is thus a genuinely far-from-equilibrium phenomenon, as claimed. 

\textbf{Discussion} Detailed experiments on colloidal particles corroborated
by analytical theory revealed a fundamental asymmetry in thermal relaxation
upon a rapid change of temperature; for thermodynamically equidistant temperature quenches as well as between two fixed temperatures,
heating is always faster than cooling. Moreover, the microscopic pathways
followed by a system during heating and cooling, respectively, are
fundamentally different. Therefore, except very near to thermodynamic
equilibrium, thermal relaxation, in general,
does \emph{not} evolve
quasistatically through quasi-equilibria even for systems with a single
energy minimum. We, therefore, witness a breakdown of the ``near
equilibrium'' paradigm of classical non-equilibrium thermodynamics
\cite{MazurGroot}.\\
\indent Namely, when the system is brought rapidly out of equilibrium, such as upon a
temperature quench, the probability density of the system cannot
follow the temperature change quasi-statically and a \emph{lag} develops between the instantaneous $P^w_i(x,t)$ and the new equilibrium $P^w_{\rm eq}(x)$ 
\cite{Vaikuntanathan_2009}. This lag, which here corresponds
to $D[P^w_i(x,t)||P^w_{\rm eq}(x)]$, is nominally smaller during
heating than during cooling. This is so because for short times heating
essentially corresponds to a free expansion \cite{Lapolla2020Faster},
which is materialized as an overshoot of statistical velocity
and is characterized by a smaller dissipated work. The latter in turn
bounds from above the maximal lag that can develop
\cite{Vaikuntanathan2009Dissipation}. Initial free expansion during
heating also
explains the faster departure from the initial equilibrium within the
backward protocol, as well as for heating and cooling between any two
fixed temperatures.    
\\
\indent Our work further underscores that there is a fundamental difference
between equidistant temperatures, thermodynamically equidistant
temperatures, and kinematically equidistant temperatures. 
The existence of a zero temperature and the $\mathrm{e}^{-1/T}$-dependence of
Boltzmann-Gibbs equilibrium statistics readily imply that raising and lowering the temperature by the same amount pushes
the system differently far from equilibrium. However, even when
temperatures are chosen to be thermodynamically equidistant, the
colder system is kinematically farther from equilibrium than the
hotter one, and yet it reaches equilibrium faster.\\
\indent Thermal relaxation, therefore, seems to be much more complex
than originally thought and our results only scratch at the
surface. Relaxation in systems with multiple energy minima
\cite{Lapolla2020Faster,Van2021Faster,Lu2017MMpemba,Kumar2022Anomalous}, time-dependent
potentials \cite{blickle2012realization,martinez2015adiabatic,martinez2016brownian,krishnamurthy2016micrometre,Koyuk,rademacher2022nonequilibrium}, driven relaxation processes \cite{martinez2016engineered,guery2022driving}, and in the presence of time-irreversible,
detailed-balance violating dynamics \cite{dieball2022direct,Polettini,Maes,Felix_2019}
still remains poorly understood, and calls for a systematic analysis
through the lens of thermal kinematics.  

\textbf{Acknowledgements} 
 This work was supported by the projects EQC2018-004693-P, PID2020-116567GB-C22, PID2021-128970OA-I00 and PID2021-127427NB-I00 funded by MCIN/AEI/ 10.13039/501100011033/ FEDER, UE, and by FEDER/Junta de Andaluc\'ia-Consejer\'ia de Transformaci\'on Econ\'omica, Industria, Conocimiento y Universidades / Proyectos P18-FR-3583 and A-FQM-644-UGR20. M.I. acknowledges financial support from Ministerio de Universidades (Spain) and Universidad de Granada for his FPU grant FPU21/02569. C.D. acknowledges financial support from the Studienstiftung des Deutschen Volkes, A.G. acknowledges financial support from the German Research Foundation (DFG) through the Emmy
Noether Program GO 2762/1-2.

\bibliography{referencias.bib}

\begin{thebibliography}{45}%
\makeatletter
\providecommand \@ifxundefined [1]{%
 \@ifx{#1\undefined}
}%
\providecommand \@ifnum [1]{%
 \ifnum #1\expandafter \@firstoftwo
 \else \expandafter \@secondoftwo
 \fi
}%
\providecommand \@ifx [1]{%
 \ifx #1\expandafter \@firstoftwo
 \else \expandafter \@secondoftwo
 \fi
}%
\providecommand \natexlab [1]{#1}%
\providecommand \enquote  [1]{``#1''}%
\providecommand \bibnamefont  [1]{#1}%
\providecommand \bibfnamefont [1]{#1}%
\providecommand \citenamefont [1]{#1}%
\providecommand \href@noop [0]{\@secondoftwo}%
\providecommand \href [0]{\begingroup \@sanitize@url \@href}%
\providecommand \@href[1]{\@@startlink{#1}\@@href}%
\providecommand \@@href[1]{\endgroup#1\@@endlink}%
\providecommand \@sanitize@url [0]{\catcode `\\12\catcode `\$12\catcode
  `\&12\catcode `\#12\catcode `\^12\catcode `\_12\catcode `\%12\relax}%
\providecommand \@@startlink[1]{}%
\providecommand \@@endlink[0]{}%
\providecommand \url  [0]{\begingroup\@sanitize@url \@url }%
\providecommand \@url [1]{\endgroup\@href {#1}{\urlprefix }}%
\providecommand \urlprefix  [0]{URL }%
\providecommand \Eprint [0]{\href }%
\providecommand \doibase [0]{https://doi.org/}%
\providecommand \selectlanguage [0]{\@gobble}%
\providecommand \bibinfo  [0]{\@secondoftwo}%
\providecommand \bibfield  [0]{\@secondoftwo}%
\providecommand \translation [1]{[#1]}%
\providecommand \BibitemOpen [0]{}%
\providecommand \bibitemStop [0]{}%
\providecommand \bibitemNoStop [0]{.\EOS\space}%
\providecommand \EOS [0]{\spacefactor3000\relax}%
\providecommand \BibitemShut  [1]{\csname bibitem#1\endcsname}%
\let\auto@bib@innerbib\@empty
\bibitem [{\citenamefont {S.~R.~de Groot}(1962)}]{MazurGroot}%
  \BibitemOpen
  \bibfield  {author} {\bibinfo {author} {\bibfnamefont {P.~M.}\ \bibnamefont
  {S.~R.~de Groot}},\ }\href@noop {} {\emph {\bibinfo {title} {Non-equilibrium
  Thermodynamics}}},\ \bibinfo {edition} {2nd}\ ed.\ (\bibinfo  {publisher}
  {North-Holland, Amsterdam},\ \bibinfo {year} {1962})\BibitemShut {NoStop}%
\bibitem [{\citenamefont {Kubo}\ \emph {et~al.}(1957)\citenamefont {Kubo},
  \citenamefont {Yokota},\ and\ \citenamefont {Nakajima}}]{Yokota}%
  \BibitemOpen
  \bibfield  {author} {\bibinfo {author} {\bibfnamefont {R.}~\bibnamefont
  {Kubo}}, \bibinfo {author} {\bibfnamefont {M.}~\bibnamefont {Yokota}},\ and\
  \bibinfo {author} {\bibfnamefont {S.}~\bibnamefont {Nakajima}},\ }\bibfield
  {title} {\bibinfo {title} {Statistical-mechanical theory of irreversible
  processes. ii. response to thermal disturbance},\ }\href
  {https://doi.org/10.1143/JPSJ.12.1203} {\bibfield  {journal} {\bibinfo
  {journal} {J. Phys. Soc. Jpn.}\ }\textbf {\bibinfo {volume} {12}},\ \bibinfo
  {pages} {1203} (\bibinfo {year} {1957})}\BibitemShut {NoStop}%
\bibitem [{\citenamefont {Lu}\ and\ \citenamefont {Raz}(2017)}]{Lu2017MMpemba}%
  \BibitemOpen
  \bibfield  {author} {\bibinfo {author} {\bibfnamefont {Z.}~\bibnamefont
  {Lu}}\ and\ \bibinfo {author} {\bibfnamefont {O.}~\bibnamefont {Raz}},\
  }\bibfield  {title} {\bibinfo {title} {Nonequilibrium thermodynamics of the
  {Markovian} {Mpemba} effect and its inverse},\ }\href
  {https://doi.org/10.1073/pnas.17012641} {\bibfield  {journal} {\bibinfo
  {journal} {Proc. Natl. Acad. Sci. U.S.A.}\ }\textbf {\bibinfo {volume}
  {114}},\ \bibinfo {pages} {5083} (\bibinfo {year} {2017})}\BibitemShut
  {NoStop}%
\bibitem [{\citenamefont {Lasanta}\ \emph {et~al.}(2017)\citenamefont
  {Lasanta}, \citenamefont {Vega~Reyes}, \citenamefont {Prados},\ and\
  \citenamefont {Santos}}]{Lasanta2017Mpemba}%
  \BibitemOpen
  \bibfield  {author} {\bibinfo {author} {\bibfnamefont {A.}~\bibnamefont
  {Lasanta}}, \bibinfo {author} {\bibfnamefont {F.}~\bibnamefont {Vega~Reyes}},
  \bibinfo {author} {\bibfnamefont {A.}~\bibnamefont {Prados}},\ and\ \bibinfo
  {author} {\bibfnamefont {A.}~\bibnamefont {Santos}},\ }\bibfield  {title}
  {\bibinfo {title} {When the hotter cools more quickly: {Mpemba} effect in
  granular fluids},\ }\href {https://doi.org/10.1103/PhysRevLett.119.148001}
  {\bibfield  {journal} {\bibinfo  {journal} {Phys. Rev. Lett.}\ }\textbf
  {\bibinfo {volume} {119}},\ \bibinfo {pages} {148001} (\bibinfo {year}
  {2017})}\BibitemShut {NoStop}%
\bibitem [{\citenamefont {Baity-Jesi}\ \emph {et~al.}(2019)\citenamefont
  {Baity-Jesi}, \citenamefont {Calore}, \citenamefont {Cruz}, \citenamefont
  {Fernandez}, \citenamefont {Gil-Narvion}, \citenamefont {Gordillo-Guerrero},
  \citenamefont {I\~niguez}, \citenamefont {Lasanta}, \citenamefont {Maiorano},
  \citenamefont {Marinari}, \citenamefont {Martin-Mayor}, \citenamefont
  {Moreno-Gordo}, \citenamefont {Mu\~noz Sudupe}, \citenamefont {Navarro},
  \citenamefont {Parisi}, \citenamefont {Perez-Gaviro}, \citenamefont
  {Ricci-Tersenghi}, \citenamefont {Ruiz-Lorenzo}, \citenamefont
  {Fabio~Schifano}, \citenamefont {Seoane}, \citenamefont {Tarancon},
  \citenamefont {Tripiccione},\ and\ \citenamefont
  {Yllanes}}]{Baity2019MpembaSG}%
  \BibitemOpen
  \bibfield  {author} {\bibinfo {author} {\bibfnamefont {M.}~\bibnamefont
  {Baity-Jesi}}, \bibinfo {author} {\bibfnamefont {E.}~\bibnamefont {Calore}},
  \bibinfo {author} {\bibfnamefont {A.}~\bibnamefont {Cruz}}, \bibinfo {author}
  {\bibfnamefont {L.~A.}\ \bibnamefont {Fernandez}}, \bibinfo {author}
  {\bibfnamefont {J.~M.}\ \bibnamefont {Gil-Narvion}}, \bibinfo {author}
  {\bibfnamefont {A.}~\bibnamefont {Gordillo-Guerrero}}, \bibinfo {author}
  {\bibfnamefont {D.}~\bibnamefont {I\~niguez}}, \bibinfo {author}
  {\bibfnamefont {A.}~\bibnamefont {Lasanta}}, \bibinfo {author} {\bibfnamefont
  {A.}~\bibnamefont {Maiorano}}, \bibinfo {author} {\bibfnamefont
  {E.}~\bibnamefont {Marinari}}, \bibinfo {author} {\bibfnamefont
  {V.}~\bibnamefont {Martin-Mayor}}, \bibinfo {author} {\bibfnamefont
  {J.}~\bibnamefont {Moreno-Gordo}}, \bibinfo {author} {\bibfnamefont
  {A.}~\bibnamefont {Mu\~noz Sudupe}}, \bibinfo {author} {\bibfnamefont
  {D.}~\bibnamefont {Navarro}}, \bibinfo {author} {\bibfnamefont
  {G.}~\bibnamefont {Parisi}}, \bibinfo {author} {\bibfnamefont
  {S.}~\bibnamefont {Perez-Gaviro}}, \bibinfo {author} {\bibfnamefont
  {F.}~\bibnamefont {Ricci-Tersenghi}}, \bibinfo {author} {\bibfnamefont
  {J.~J.}\ \bibnamefont {Ruiz-Lorenzo}}, \bibinfo {author} {\bibfnamefont
  {S.}~\bibnamefont {Fabio~Schifano}}, \bibinfo {author} {\bibfnamefont
  {B.}~\bibnamefont {Seoane}}, \bibinfo {author} {\bibfnamefont
  {A.}~\bibnamefont {Tarancon}}, \bibinfo {author} {\bibfnamefont
  {R.}~\bibnamefont {Tripiccione}},\ and\ \bibinfo {author} {\bibfnamefont
  {D.}~\bibnamefont {Yllanes}},\ }\bibfield  {title} {\bibinfo {title} {The
  {Mpemba} effect in spin glasses is a persistent memory effect},\ }\href
  {https://doi.org/10.1073/pnas.18198031} {\bibfield  {journal} {\bibinfo
  {journal} {Proc. Natl. Acad. Sci. U.S.A.}\ }\textbf {\bibinfo {volume}
  {116}},\ \bibinfo {pages} {15350} (\bibinfo {year} {2019})}\BibitemShut
  {NoStop}%
\bibitem [{\citenamefont {Kumar}\ and\ \citenamefont
  {Bechhoefer}(2020)}]{Kumar2020EMpemba}%
  \BibitemOpen
  \bibfield  {author} {\bibinfo {author} {\bibfnamefont {A.}~\bibnamefont
  {Kumar}}\ and\ \bibinfo {author} {\bibfnamefont {J.}~\bibnamefont
  {Bechhoefer}},\ }\bibfield  {title} {\bibinfo {title} {Exponentially faster
  cooling in a colloidal system},\ }\href
  {https://doi.org/10.1038/s41586-020-2560-x} {\bibfield  {journal} {\bibinfo
  {journal} {Nature}\ }\textbf {\bibinfo {volume} {584}},\ \bibinfo {pages}
  {64} (\bibinfo {year} {2020})}\BibitemShut {NoStop}%
\bibitem [{\citenamefont {Carollo}\ \emph {et~al.}(2021)\citenamefont
  {Carollo}, \citenamefont {Lasanta},\ and\ \citenamefont
  {Lesanovsky}}]{Carollo2021QMpemba}%
  \BibitemOpen
  \bibfield  {author} {\bibinfo {author} {\bibfnamefont {F.}~\bibnamefont
  {Carollo}}, \bibinfo {author} {\bibfnamefont {A.}~\bibnamefont {Lasanta}},\
  and\ \bibinfo {author} {\bibfnamefont {I.}~\bibnamefont {Lesanovsky}},\
  }\bibfield  {title} {\bibinfo {title} {Exponentially accelerated approach to
  stationarity in {Markovian} open quantum systems through the {Mpemba}
  effect},\ }\href {https://doi.org/10.1103/PhysRevLett.127.060401} {\bibfield
  {journal} {\bibinfo  {journal} {Phys. Rev. Lett.}\ }\textbf {\bibinfo
  {volume} {127}},\ \bibinfo {pages} {060401} (\bibinfo {year}
  {2021})}\BibitemShut {NoStop}%
\bibitem [{\citenamefont {Kumar}\ \emph {et~al.}(2022)\citenamefont {Kumar},
  \citenamefont {Chétrite},\ and\ \citenamefont
  {Bechhoefer}}]{Kumar2022Anomalous}%
  \BibitemOpen
  \bibfield  {author} {\bibinfo {author} {\bibfnamefont {A.}~\bibnamefont
  {Kumar}}, \bibinfo {author} {\bibfnamefont {R.}~\bibnamefont {Chétrite}},\
  and\ \bibinfo {author} {\bibfnamefont {J.}~\bibnamefont {Bechhoefer}},\
  }\bibfield  {title} {\bibinfo {title} {Anomalous heating in a colloidal
  system},\ }\bibfield  {journal} {\bibinfo  {journal} {Proc. Natl. Acad. Sci.
  U.S.A.}\ }\textbf {\bibinfo {volume} {119}},\ \href
  {https://doi.org/10.1073/pnas.2118484119} {10.1073/pnas.2118484119} (\bibinfo
  {year} {2022})\BibitemShut {NoStop}%
\bibitem [{\citenamefont {Klich}\ \emph {et~al.}(2019)\citenamefont {Klich},
  \citenamefont {Raz}, \citenamefont {Hirschberg},\ and\ \citenamefont
  {Vucelja}}]{Klich}%
  \BibitemOpen
  \bibfield  {author} {\bibinfo {author} {\bibfnamefont {I.}~\bibnamefont
  {Klich}}, \bibinfo {author} {\bibfnamefont {O.}~\bibnamefont {Raz}}, \bibinfo
  {author} {\bibfnamefont {O.}~\bibnamefont {Hirschberg}},\ and\ \bibinfo
  {author} {\bibfnamefont {M.}~\bibnamefont {Vucelja}},\ }\bibfield  {title}
  {\bibinfo {title} {Mpemba index and anomalous relaxation},\ }\href
  {https://doi.org/10.1103/PhysRevX.9.021060} {\bibfield  {journal} {\bibinfo
  {journal} {Phys. Rev. X}\ }\textbf {\bibinfo {volume} {8}},\ \bibinfo {pages}
  {021060} (\bibinfo {year} {2019})}\BibitemShut {NoStop}%
\bibitem [{\citenamefont {Josserand}\ \emph {et~al.}(2000)\citenamefont
  {Josserand}, \citenamefont {Tkachenko}, \citenamefont {Mueth},\ and\
  \citenamefont {Jaeger}}]{Kovacs_granular}%
  \BibitemOpen
  \bibfield  {author} {\bibinfo {author} {\bibfnamefont {C.}~\bibnamefont
  {Josserand}}, \bibinfo {author} {\bibfnamefont {A.~V.}\ \bibnamefont
  {Tkachenko}}, \bibinfo {author} {\bibfnamefont {D.~M.}\ \bibnamefont
  {Mueth}},\ and\ \bibinfo {author} {\bibfnamefont {H.~M.}\ \bibnamefont
  {Jaeger}},\ }\bibfield  {title} {\bibinfo {title} {Memory effects in granular
  materials},\ }\href {https://doi.org/10.1103/PhysRevLett.85.3632} {\bibfield
  {journal} {\bibinfo  {journal} {Phys. Rev. Lett.}\ }\textbf {\bibinfo
  {volume} {85}},\ \bibinfo {pages} {3632} (\bibinfo {year}
  {2000})}\BibitemShut {NoStop}%
\bibitem [{\citenamefont {Lahini}\ \emph {et~al.}(2017)\citenamefont {Lahini},
  \citenamefont {Gottesman}, \citenamefont {Amir},\ and\ \citenamefont
  {Rubinstein}}]{Kovacs_disordered}%
  \BibitemOpen
  \bibfield  {author} {\bibinfo {author} {\bibfnamefont {Y.}~\bibnamefont
  {Lahini}}, \bibinfo {author} {\bibfnamefont {O.}~\bibnamefont {Gottesman}},
  \bibinfo {author} {\bibfnamefont {A.}~\bibnamefont {Amir}},\ and\ \bibinfo
  {author} {\bibfnamefont {S.~M.}\ \bibnamefont {Rubinstein}},\ }\bibfield
  {title} {\bibinfo {title} {Nonmonotonic aging and memory retention in
  disordered mechanical systems},\ }\href
  {https://doi.org/10.1103/PhysRevLett.118.085501} {\bibfield  {journal}
  {\bibinfo  {journal} {Phys. Rev. Lett.}\ }\textbf {\bibinfo {volume} {118}},\
  \bibinfo {pages} {085501} (\bibinfo {year} {2017})}\BibitemShut {NoStop}%
\bibitem [{\citenamefont {Morgan}\ \emph {et~al.}(2020)\citenamefont {Morgan},
  \citenamefont {Avinery}, \citenamefont {Rahamim}, \citenamefont {Beck},\ and\
  \citenamefont {Saleh}}]{Kovacs_protein}%
  \BibitemOpen
  \bibfield  {author} {\bibinfo {author} {\bibfnamefont {I.~L.}\ \bibnamefont
  {Morgan}}, \bibinfo {author} {\bibfnamefont {R.}~\bibnamefont {Avinery}},
  \bibinfo {author} {\bibfnamefont {G.}~\bibnamefont {Rahamim}}, \bibinfo
  {author} {\bibfnamefont {R.}~\bibnamefont {Beck}},\ and\ \bibinfo {author}
  {\bibfnamefont {O.~A.}\ \bibnamefont {Saleh}},\ }\bibfield  {title} {\bibinfo
  {title} {Glassy dynamics and memory effects in an intrinsically disordered
  protein construct},\ }\href {https://doi.org/10.1103/PhysRevLett.125.058001}
  {\bibfield  {journal} {\bibinfo  {journal} {Phys. Rev. Lett.}\ }\textbf
  {\bibinfo {volume} {125}},\ \bibinfo {pages} {058001} (\bibinfo {year}
  {2020})}\BibitemShut {NoStop}%
\bibitem [{\citenamefont {Militaru}\ \emph {et~al.}(2021)\citenamefont
  {Militaru}, \citenamefont {Lasanta}, \citenamefont {Frimmer}, \citenamefont
  {Bonilla}, \citenamefont {Novotny},\ and\ \citenamefont
  {Rica}}]{Militaru2021Kovacs}%
  \BibitemOpen
  \bibfield  {author} {\bibinfo {author} {\bibfnamefont {A.}~\bibnamefont
  {Militaru}}, \bibinfo {author} {\bibfnamefont {A.}~\bibnamefont {Lasanta}},
  \bibinfo {author} {\bibfnamefont {M.}~\bibnamefont {Frimmer}}, \bibinfo
  {author} {\bibfnamefont {L.~L.}\ \bibnamefont {Bonilla}}, \bibinfo {author}
  {\bibfnamefont {L.}~\bibnamefont {Novotny}},\ and\ \bibinfo {author}
  {\bibfnamefont {R.~A.}\ \bibnamefont {Rica}},\ }\bibfield  {title} {\bibinfo
  {title} {Kovacs memory effect with an optically levitated nanoparticle},\
  }\href {https://doi.org/10.1103/PhysRevLett.127.130603} {\bibfield  {journal}
  {\bibinfo  {journal} {Phys. Rev. Lett.}\ }\textbf {\bibinfo {volume} {127}},\
  \bibinfo {pages} {130603} (\bibinfo {year} {2021})}\BibitemShut {NoStop}%
\bibitem [{\citenamefont {Riechers}\ \emph {et~al.}(2022)\citenamefont
  {Riechers}, \citenamefont {Roed}, \citenamefont {Mehri}, \citenamefont
  {Ingebrigtsen}, \citenamefont {Hecksher}, \citenamefont {Dyre},\ and\
  \citenamefont {Niss}}]{Kovacs_liquid}%
  \BibitemOpen
  \bibfield  {author} {\bibinfo {author} {\bibfnamefont {B.}~\bibnamefont
  {Riechers}}, \bibinfo {author} {\bibfnamefont {L.~A.}\ \bibnamefont {Roed}},
  \bibinfo {author} {\bibfnamefont {S.}~\bibnamefont {Mehri}}, \bibinfo
  {author} {\bibfnamefont {T.~S.}\ \bibnamefont {Ingebrigtsen}}, \bibinfo
  {author} {\bibfnamefont {T.}~\bibnamefont {Hecksher}}, \bibinfo {author}
  {\bibfnamefont {J.~C.}\ \bibnamefont {Dyre}},\ and\ \bibinfo {author}
  {\bibfnamefont {K.}~\bibnamefont {Niss}},\ }\bibfield  {title} {\bibinfo
  {title} {Predicting nonlinear physical aging of glasses from equilibrium
  relaxation via the material time},\ }\href
  {https://doi.org/10.1126/sciadv.abl9809} {\bibfield  {journal} {\bibinfo
  {journal} {Sci. Adv.}\ }\textbf {\bibinfo {volume} {8}},\ \bibinfo {pages}
  {eabl9809} (\bibinfo {year} {2022})},\ \Eprint
  {https://arxiv.org/abs/https://www.science.org/doi/pdf/10.1126/sciadv.abl9809}
  {https://www.science.org/doi/pdf/10.1126/sciadv.abl9809} \BibitemShut
  {NoStop}%
\bibitem [{\citenamefont {Lapolla}\ and\ \citenamefont
  {Godec}(2020)}]{Lapolla2020Faster}%
  \BibitemOpen
  \bibfield  {author} {\bibinfo {author} {\bibfnamefont {A.}~\bibnamefont
  {Lapolla}}\ and\ \bibinfo {author} {\bibfnamefont {A.}~\bibnamefont
  {Godec}},\ }\bibfield  {title} {\bibinfo {title} {Faster uphill relaxation in
  thermodynamically equidistant temperature quenches},\ }\bibfield  {journal}
  {\bibinfo  {journal} {Phys. Rev. Lett.}\ }\textbf {\bibinfo {volume} {125}},\
  \href {https://doi.org/10.1103/PHYSREVLETT.125.110602}
  {10.1103/PHYSREVLETT.125.110602} (\bibinfo {year} {2020})\BibitemShut
  {NoStop}%
\bibitem [{\citenamefont {Van~Vu}\ and\ \citenamefont
  {Hasegawa}(2021)}]{Van2021Faster}%
  \BibitemOpen
  \bibfield  {author} {\bibinfo {author} {\bibfnamefont {T.}~\bibnamefont
  {Van~Vu}}\ and\ \bibinfo {author} {\bibfnamefont {T.}~\bibnamefont
  {Hasegawa}},\ }\bibfield  {title} {\bibinfo {title} {Toward relaxation
  asymmetry: Heating is faster than cooling},\ }\href
  {https://doi.org/10.1103/PhysRevResearch.3.043160} {\bibfield  {journal}
  {\bibinfo  {journal} {Phys. Rev. Res.}\ }\textbf {\bibinfo {volume} {3}},\
  \bibinfo {pages} {043160} (\bibinfo {year} {2021})}\BibitemShut {NoStop}%
\bibitem [{\citenamefont {Meibohm}\ \emph {et~al.}(2021)\citenamefont
  {Meibohm}, \citenamefont {Forastiere}, \citenamefont {Adeleke-Larodo},\ and\
  \citenamefont {Proesmans}}]{meibohm2021relaxation}%
  \BibitemOpen
  \bibfield  {author} {\bibinfo {author} {\bibfnamefont {J.}~\bibnamefont
  {Meibohm}}, \bibinfo {author} {\bibfnamefont {D.}~\bibnamefont {Forastiere}},
  \bibinfo {author} {\bibfnamefont {T.}~\bibnamefont {Adeleke-Larodo}},\ and\
  \bibinfo {author} {\bibfnamefont {K.}~\bibnamefont {Proesmans}},\ }\bibfield
  {title} {\bibinfo {title} {Relaxation-speed crossover in anharmonic
  potentials},\ }\href@noop {} {\bibfield  {journal} {\bibinfo  {journal}
  {Phys. Rev. E}\ }\textbf {\bibinfo {volume} {104}},\ \bibinfo {pages}
  {L032105} (\bibinfo {year} {2021})}\BibitemShut {NoStop}%
\bibitem [{\citenamefont {Seifert}(2012{\natexlab{a}})}]{Udo_2012}%
  \BibitemOpen
  \bibfield  {author} {\bibinfo {author} {\bibfnamefont {U.}~\bibnamefont
  {Seifert}},\ }\bibfield  {title} {\bibinfo {title} {Stochastic
  thermodynamics, fluctuation theorems and molecular machines},\ }\href
  {https://doi.org/10.1088/0034-4885/75/12/126001} {\bibfield  {journal}
  {\bibinfo  {journal} {Rep. Prog. Phys.}\ }\textbf {\bibinfo {volume} {75}},\
  \bibinfo {pages} {126001} (\bibinfo {year} {2012}{\natexlab{a}})}\BibitemShut
  {NoStop}%
\bibitem [{\citenamefont {Sekimoto}(2010)}]{Sekimoto2010}%
  \BibitemOpen
  \bibfield  {author} {\bibinfo {author} {\bibfnamefont {K.}~\bibnamefont
  {Sekimoto}},\ }\href@noop {} {\emph {\bibinfo {title} {Stochastic
  Energetics}}},\ \bibinfo {edition} {1st}\ ed.\ (\bibinfo  {publisher}
  {Springer Lecture Notes in Physics},\ \bibinfo {year} {2010})\BibitemShut
  {NoStop}%
\bibitem [{\citenamefont {Seifert}(2019)}]{Udo_2019}%
  \BibitemOpen
  \bibfield  {author} {\bibinfo {author} {\bibfnamefont {U.}~\bibnamefont
  {Seifert}},\ }\bibfield  {title} {\bibinfo {title} {From stochastic
  thermodynamics to thermodynamic inference},\ }\href
  {https://doi.org/10.1146/annurev-conmatphys-031218-013554} {\bibfield
  {journal} {\bibinfo  {journal} {Annu. Rev. Condens. Mat. Phys.}\ }\textbf
  {\bibinfo {volume} {10}},\ \bibinfo {pages} {171–192} (\bibinfo {year}
  {2019})}\BibitemShut {NoStop}%
\bibitem [{\citenamefont
  {Seifert}(2012{\natexlab{b}})}]{Seifert2012Stochastic}%
  \BibitemOpen
  \bibfield  {author} {\bibinfo {author} {\bibfnamefont {U.}~\bibnamefont
  {Seifert}},\ }\bibfield  {title} {\bibinfo {title} {Stochastic
  thermodynamics, fluctuation theorems and molecular machines},\ }\bibfield
  {journal} {\bibinfo  {journal} {Rep. Prog. Phys.}\ }\textbf {\bibinfo
  {volume} {75}},\ \href {https://doi.org/10.1088/0034-4885/75/12/126001}
  {10.1088/0034-4885/75/12/126001} (\bibinfo {year}
  {2012}{\natexlab{b}})\BibitemShut {NoStop}%
\bibitem [{\citenamefont {Blickle}\ and\ \citenamefont
  {Bechinger}(2012)}]{blickle2012realization}%
  \BibitemOpen
  \bibfield  {author} {\bibinfo {author} {\bibfnamefont {V.}~\bibnamefont
  {Blickle}}\ and\ \bibinfo {author} {\bibfnamefont {C.}~\bibnamefont
  {Bechinger}},\ }\bibfield  {title} {\bibinfo {title} {Realization of a
  micrometre-sized stochastic heat engine},\ }\href@noop {} {\bibfield
  {journal} {\bibinfo  {journal} {Nat. Phys.}\ }\textbf {\bibinfo {volume}
  {8}},\ \bibinfo {pages} {143} (\bibinfo {year} {2012})}\BibitemShut {NoStop}%
\bibitem [{\citenamefont {Mart{\'\i}nez}\ \emph {et~al.}(2015)\citenamefont
  {Mart{\'\i}nez}, \citenamefont {Rold{\'a}n}, \citenamefont {Dinis},
  \citenamefont {Petrov},\ and\ \citenamefont {Rica}}]{martinez2015adiabatic}%
  \BibitemOpen
  \bibfield  {author} {\bibinfo {author} {\bibfnamefont {I.~A.}\ \bibnamefont
  {Mart{\'\i}nez}}, \bibinfo {author} {\bibfnamefont {{\'E}.}~\bibnamefont
  {Rold{\'a}n}}, \bibinfo {author} {\bibfnamefont {L.}~\bibnamefont {Dinis}},
  \bibinfo {author} {\bibfnamefont {D.}~\bibnamefont {Petrov}},\ and\ \bibinfo
  {author} {\bibfnamefont {R.~A.}\ \bibnamefont {Rica}},\ }\bibfield  {title}
  {\bibinfo {title} {Adiabatic processes realized with a trapped {Brownian}
  particle},\ }\href@noop {} {\bibfield  {journal} {\bibinfo  {journal} {Phys.
  Rev. Lett.}\ }\textbf {\bibinfo {volume} {114}},\ \bibinfo {pages} {120601}
  (\bibinfo {year} {2015})}\BibitemShut {NoStop}%
\bibitem [{\citenamefont {Lebowitz}\ and\ \citenamefont
  {Bergmann}(1957)}]{LEBOWITZ19571}%
  \BibitemOpen
  \bibfield  {author} {\bibinfo {author} {\bibfnamefont {J.~L.}\ \bibnamefont
  {Lebowitz}}\ and\ \bibinfo {author} {\bibfnamefont {P.~G.}\ \bibnamefont
  {Bergmann}},\ }\bibfield  {title} {\bibinfo {title} {Irreversible gibbsian
  ensembles},\ }\href
  {https://doi.org/https://doi.org/10.1016/0003-4916(57)90002-7} {\bibfield
  {journal} {\bibinfo  {journal} {Ann. Phys.}\ }\textbf {\bibinfo {volume}
  {1}},\ \bibinfo {pages} {1} (\bibinfo {year} {1957})}\BibitemShut {NoStop}%
\bibitem [{\citenamefont {Vaikuntanathan}\ and\ \citenamefont
  {Jarzynski}(2009{\natexlab{a}})}]{Vaikuntanathan_2009}%
  \BibitemOpen
  \bibfield  {author} {\bibinfo {author} {\bibfnamefont {S.}~\bibnamefont
  {Vaikuntanathan}}\ and\ \bibinfo {author} {\bibfnamefont {C.}~\bibnamefont
  {Jarzynski}},\ }\bibfield  {title} {\bibinfo {title} {Dissipation and lag in
  irreversible processes},\ }\href {https://doi.org/10.1209/0295-5075/87/60005}
  {\bibfield  {journal} {\bibinfo  {journal} {Europhys. Lett.}\ }\textbf
  {\bibinfo {volume} {87}},\ \bibinfo {pages} {60005} (\bibinfo {year}
  {2009}{\natexlab{a}})}\BibitemShut {NoStop}%
\bibitem [{\citenamefont {Shiraishi}\ and\ \citenamefont
  {Saito}(2019)}]{shiraishi2019information}%
  \BibitemOpen
  \bibfield  {author} {\bibinfo {author} {\bibfnamefont {N.}~\bibnamefont
  {Shiraishi}}\ and\ \bibinfo {author} {\bibfnamefont {K.}~\bibnamefont
  {Saito}},\ }\bibfield  {title} {\bibinfo {title} {Information-theoretical
  bound of the irreversibility in thermal relaxation processes},\ }\href@noop
  {} {\bibfield  {journal} {\bibinfo  {journal} {Phys. Rev. Lett.}\ }\textbf
  {\bibinfo {volume} {123}},\ \bibinfo {pages} {110603} (\bibinfo {year}
  {2019})}\BibitemShut {NoStop}%
\bibitem [{\citenamefont {Ito}\ and\ \citenamefont
  {Dechant}(2020)}]{Ito2020Stochastic}%
  \BibitemOpen
  \bibfield  {author} {\bibinfo {author} {\bibfnamefont {S.}~\bibnamefont
  {Ito}}\ and\ \bibinfo {author} {\bibfnamefont {A.}~\bibnamefont {Dechant}},\
  }\bibfield  {title} {\bibinfo {title} {Stochastic time evolution, information
  geometry, and the {C}ramér-{Rao} bound},\ }\bibfield  {journal} {\bibinfo
  {journal} {Phys. Rev. X}\ }\textbf {\bibinfo {volume} {10}},\ \href
  {https://doi.org/10.1103/PhysRevX.10.021056} {10.1103/PhysRevX.10.021056}
  (\bibinfo {year} {2020})\BibitemShut {NoStop}%
\bibitem [{\citenamefont {Barato}\ and\ \citenamefont
  {Seifert}(2015)}]{barato2015thermodynamic}%
  \BibitemOpen
  \bibfield  {author} {\bibinfo {author} {\bibfnamefont {A.~C.}\ \bibnamefont
  {Barato}}\ and\ \bibinfo {author} {\bibfnamefont {U.}~\bibnamefont
  {Seifert}},\ }\bibfield  {title} {\bibinfo {title} {Thermodynamic uncertainty
  relation for biomolecular processes},\ }\href@noop {} {\bibfield  {journal}
  {\bibinfo  {journal} {Phys. Rev. Lett.}\ }\textbf {\bibinfo {volume} {114}},\
  \bibinfo {pages} {158101} (\bibinfo {year} {2015})}\BibitemShut {NoStop}%
\bibitem [{\citenamefont {Gingrich}\ \emph {et~al.}(2016)\citenamefont
  {Gingrich}, \citenamefont {Horowitz}, \citenamefont {Perunov},\ and\
  \citenamefont {England}}]{gingrich2016dissipation}%
  \BibitemOpen
  \bibfield  {author} {\bibinfo {author} {\bibfnamefont {T.~R.}\ \bibnamefont
  {Gingrich}}, \bibinfo {author} {\bibfnamefont {J.~M.}\ \bibnamefont
  {Horowitz}}, \bibinfo {author} {\bibfnamefont {N.}~\bibnamefont {Perunov}},\
  and\ \bibinfo {author} {\bibfnamefont {J.~L.}\ \bibnamefont {England}},\
  }\bibfield  {title} {\bibinfo {title} {Dissipation bounds all steady-state
  current fluctuations},\ }\href@noop {} {\bibfield  {journal} {\bibinfo
  {journal} {Phys. Rev. Lett.}\ }\textbf {\bibinfo {volume} {116}},\ \bibinfo
  {pages} {120601} (\bibinfo {year} {2016})}\BibitemShut {NoStop}%
\bibitem [{\citenamefont {Dechant}\ and\ \citenamefont
  {Sasa}(2021)}]{Dechant_PRX}%
  \BibitemOpen
  \bibfield  {author} {\bibinfo {author} {\bibfnamefont {A.}~\bibnamefont
  {Dechant}}\ and\ \bibinfo {author} {\bibfnamefont {S.-i.}\ \bibnamefont
  {Sasa}},\ }\bibfield  {title} {\bibinfo {title} {Improving thermodynamic
  bounds using correlations},\ }\href
  {https://doi.org/10.1103/PhysRevX.11.041061} {\bibfield  {journal} {\bibinfo
  {journal} {Phys. Rev. X}\ }\textbf {\bibinfo {volume} {11}},\ \bibinfo
  {pages} {041061} (\bibinfo {year} {2021})}\BibitemShut {NoStop}%
\bibitem [{\citenamefont {Dieball}\ and\ \citenamefont
  {Godec}(2022)}]{dieball2022direct}%
  \BibitemOpen
  \bibfield  {author} {\bibinfo {author} {\bibfnamefont {C.}~\bibnamefont
  {Dieball}}\ and\ \bibinfo {author} {\bibfnamefont {A.}~\bibnamefont
  {Godec}},\ }\bibfield  {title} {\bibinfo {title} {Direct route to
  thermodynamic uncertainty relations and their saturation},\ }\bibfield
  {journal} {\bibinfo  {journal} {Phys. Rev. Lett.}\ }\textbf {\bibinfo
  {volume} {\text{article in press (arXiv:2208.06402)}}},\ \href
  {https://doi.org/10.48550/arXiv.2208.06402} {10.48550/arXiv.2208.06402}
  (\bibinfo {year} {2022})\BibitemShut {NoStop}%
\bibitem [{\citenamefont {Okuyama}\ and\ \citenamefont
  {Ohzeki}(2018)}]{okuyama2018quantum}%
  \BibitemOpen
  \bibfield  {author} {\bibinfo {author} {\bibfnamefont {M.}~\bibnamefont
  {Okuyama}}\ and\ \bibinfo {author} {\bibfnamefont {M.}~\bibnamefont
  {Ohzeki}},\ }\bibfield  {title} {\bibinfo {title} {Quantum speed limit is not
  quantum},\ }\href@noop {} {\bibfield  {journal} {\bibinfo  {journal} {Phys.
  Rev. Lett.}\ }\textbf {\bibinfo {volume} {120}},\ \bibinfo {pages} {070402}
  (\bibinfo {year} {2018})}\BibitemShut {NoStop}%
\bibitem [{\citenamefont {Shiraishi}\ \emph {et~al.}(2018)\citenamefont
  {Shiraishi}, \citenamefont {Funo},\ and\ \citenamefont
  {Saito}}]{shiraishi2018speed}%
  \BibitemOpen
  \bibfield  {author} {\bibinfo {author} {\bibfnamefont {N.}~\bibnamefont
  {Shiraishi}}, \bibinfo {author} {\bibfnamefont {K.}~\bibnamefont {Funo}},\
  and\ \bibinfo {author} {\bibfnamefont {K.}~\bibnamefont {Saito}},\ }\bibfield
   {title} {\bibinfo {title} {Speed limit for classical stochastic processes},\
  }\href@noop {} {\bibfield  {journal} {\bibinfo  {journal} {Phys. Rev. Lett.}\
  }\textbf {\bibinfo {volume} {121}},\ \bibinfo {pages} {070601} (\bibinfo
  {year} {2018})}\BibitemShut {NoStop}%
\bibitem [{\citenamefont {Crooks}(2007)}]{Crooks2007Measuring}%
  \BibitemOpen
  \bibfield  {author} {\bibinfo {author} {\bibfnamefont {G.~E.}\ \bibnamefont
  {Crooks}},\ }\bibfield  {title} {\bibinfo {title} {Measuring thermodynamic
  length},\ }\href {https://doi.org/10.1103/PhysRevLett.99.100602} {\bibfield
  {journal} {\bibinfo  {journal} {Phys. Rev. Lett.}\ }\textbf {\bibinfo
  {volume} {99}},\ \bibinfo {pages} {100602} (\bibinfo {year}
  {2007})}\BibitemShut {NoStop}%
\bibitem [{\citenamefont {Cover}\ and\ \citenamefont
  {Thomas}(2006)}]{Cover2006Elements}%
  \BibitemOpen
  \bibfield  {author} {\bibinfo {author} {\bibfnamefont {T.}~\bibnamefont
  {Cover}}\ and\ \bibinfo {author} {\bibfnamefont {J.}~\bibnamefont {Thomas}},\
  }\href@noop {} {\emph {\bibinfo {title} {Elements of Information Theory}}},\
  \bibinfo {edition} {2nd}\ ed.\ (\bibinfo  {publisher} {Wiley, Hoboken, New
  Jersey},\ \bibinfo {year} {2006})\BibitemShut {NoStop}%
\bibitem [{\citenamefont {Vaikuntanathan}\ and\ \citenamefont
  {Jarzynski}(2009{\natexlab{b}})}]{Vaikuntanathan2009Dissipation}%
  \BibitemOpen
  \bibfield  {author} {\bibinfo {author} {\bibfnamefont {S.}~\bibnamefont
  {Vaikuntanathan}}\ and\ \bibinfo {author} {\bibfnamefont {C.}~\bibnamefont
  {Jarzynski}},\ }\bibfield  {title} {\bibinfo {title} {Dissipation and lag in
  irreversible processes},\ }\href {https://doi.org/10.1209/0295-5075/87/60005}
  {\bibfield  {journal} {\bibinfo  {journal} {Europhys. Lett.}\ }\textbf
  {\bibinfo {volume} {87}},\ \bibinfo {pages} {60005} (\bibinfo {year}
  {2009}{\natexlab{b}})}\BibitemShut {NoStop}%
\bibitem [{\citenamefont {Mart{\'\i}nez}\ \emph
  {et~al.}(2016{\natexlab{a}})\citenamefont {Mart{\'\i}nez}, \citenamefont
  {Rold{\'a}n}, \citenamefont {Dinis}, \citenamefont {Petrov}, \citenamefont
  {Parrondo},\ and\ \citenamefont {Rica}}]{martinez2016brownian}%
  \BibitemOpen
  \bibfield  {author} {\bibinfo {author} {\bibfnamefont {I.~A.}\ \bibnamefont
  {Mart{\'\i}nez}}, \bibinfo {author} {\bibfnamefont {{\'E}.}~\bibnamefont
  {Rold{\'a}n}}, \bibinfo {author} {\bibfnamefont {L.}~\bibnamefont {Dinis}},
  \bibinfo {author} {\bibfnamefont {D.}~\bibnamefont {Petrov}}, \bibinfo
  {author} {\bibfnamefont {J.~M.}\ \bibnamefont {Parrondo}},\ and\ \bibinfo
  {author} {\bibfnamefont {R.~A.}\ \bibnamefont {Rica}},\ }\bibfield  {title}
  {\bibinfo {title} {Brownian carnot engine},\ }\href@noop {} {\bibfield
  {journal} {\bibinfo  {journal} {Nat. Phys.}\ }\textbf {\bibinfo {volume}
  {12}},\ \bibinfo {pages} {67} (\bibinfo {year}
  {2016}{\natexlab{a}})}\BibitemShut {NoStop}%
\bibitem [{\citenamefont {Krishnamurthy}\ \emph {et~al.}(2016)\citenamefont
  {Krishnamurthy}, \citenamefont {Ghosh}, \citenamefont {Chatterji},
  \citenamefont {Ganapathy},\ and\ \citenamefont
  {Sood}}]{krishnamurthy2016micrometre}%
  \BibitemOpen
  \bibfield  {author} {\bibinfo {author} {\bibfnamefont {S.}~\bibnamefont
  {Krishnamurthy}}, \bibinfo {author} {\bibfnamefont {S.}~\bibnamefont
  {Ghosh}}, \bibinfo {author} {\bibfnamefont {D.}~\bibnamefont {Chatterji}},
  \bibinfo {author} {\bibfnamefont {R.}~\bibnamefont {Ganapathy}},\ and\
  \bibinfo {author} {\bibfnamefont {A.}~\bibnamefont {Sood}},\ }\bibfield
  {title} {\bibinfo {title} {A micrometre-sized heat engine operating between
  bacterial reservoirs},\ }\href@noop {} {\bibfield  {journal} {\bibinfo
  {journal} {Nat. Phys.}\ }\textbf {\bibinfo {volume} {12}},\ \bibinfo {pages}
  {1134} (\bibinfo {year} {2016})}\BibitemShut {NoStop}%
\bibitem [{\citenamefont {Koyuk}\ and\ \citenamefont {Seifert}(2020)}]{Koyuk}%
  \BibitemOpen
  \bibfield  {author} {\bibinfo {author} {\bibfnamefont {T.}~\bibnamefont
  {Koyuk}}\ and\ \bibinfo {author} {\bibfnamefont {U.}~\bibnamefont
  {Seifert}},\ }\bibfield  {title} {\bibinfo {title} {Thermodynamic uncertainty
  relation for time-dependent driving},\ }\href
  {https://doi.org/10.1103/PhysRevLett.125.260604} {\bibfield  {journal}
  {\bibinfo  {journal} {Phys. Rev. Lett.}\ }\textbf {\bibinfo {volume} {125}},\
  \bibinfo {pages} {260604} (\bibinfo {year} {2020})}\BibitemShut {NoStop}%
\bibitem [{\citenamefont {Rademacher}\ \emph {et~al.}(2022)\citenamefont
  {Rademacher}, \citenamefont {Konopik}, \citenamefont {Debiossac},
  \citenamefont {Grass}, \citenamefont {Lutz},\ and\ \citenamefont
  {Kiesel}}]{rademacher2022nonequilibrium}%
  \BibitemOpen
  \bibfield  {author} {\bibinfo {author} {\bibfnamefont {M.}~\bibnamefont
  {Rademacher}}, \bibinfo {author} {\bibfnamefont {M.}~\bibnamefont {Konopik}},
  \bibinfo {author} {\bibfnamefont {M.}~\bibnamefont {Debiossac}}, \bibinfo
  {author} {\bibfnamefont {D.}~\bibnamefont {Grass}}, \bibinfo {author}
  {\bibfnamefont {E.}~\bibnamefont {Lutz}},\ and\ \bibinfo {author}
  {\bibfnamefont {N.}~\bibnamefont {Kiesel}},\ }\bibfield  {title} {\bibinfo
  {title} {Nonequilibrium control of thermal and mechanical changes in a
  levitated system},\ }\bibfield  {journal} {\bibinfo  {journal} {Phys. Rev.
  Lett.}\ }\textbf {\bibinfo {volume} {128}},\ \href
  {https://doi.org/10.1103/physrevlett.128.070601}
  {10.1103/physrevlett.128.070601} (\bibinfo {year} {2022})\BibitemShut
  {NoStop}%
\bibitem [{\citenamefont {Mart{\'\i}nez}\ \emph
  {et~al.}(2016{\natexlab{b}})\citenamefont {Mart{\'\i}nez}, \citenamefont
  {Petrosyan}, \citenamefont {Gu{\'e}ry-Odelin}, \citenamefont {Trizac},\ and\
  \citenamefont {Ciliberto}}]{martinez2016engineered}%
  \BibitemOpen
  \bibfield  {author} {\bibinfo {author} {\bibfnamefont {I.~A.}\ \bibnamefont
  {Mart{\'\i}nez}}, \bibinfo {author} {\bibfnamefont {A.}~\bibnamefont
  {Petrosyan}}, \bibinfo {author} {\bibfnamefont {D.}~\bibnamefont
  {Gu{\'e}ry-Odelin}}, \bibinfo {author} {\bibfnamefont {E.}~\bibnamefont
  {Trizac}},\ and\ \bibinfo {author} {\bibfnamefont {S.}~\bibnamefont
  {Ciliberto}},\ }\bibfield  {title} {\bibinfo {title} {Engineered swift
  equilibration of a {Brownian} particle},\ }\href@noop {} {\bibfield
  {journal} {\bibinfo  {journal} {Nat. Phys.}\ }\textbf {\bibinfo {volume}
  {12}},\ \bibinfo {pages} {843} (\bibinfo {year}
  {2016}{\natexlab{b}})}\BibitemShut {NoStop}%
\bibitem [{\citenamefont {Gu{\'e}ry-Odelin}\ \emph {et~al.}(2022)\citenamefont
  {Gu{\'e}ry-Odelin}, \citenamefont {Jarzynski}, \citenamefont {Plata},
  \citenamefont {Prados},\ and\ \citenamefont {Trizac}}]{guery2022driving}%
  \BibitemOpen
  \bibfield  {author} {\bibinfo {author} {\bibfnamefont {D.}~\bibnamefont
  {Gu{\'e}ry-Odelin}}, \bibinfo {author} {\bibfnamefont {C.}~\bibnamefont
  {Jarzynski}}, \bibinfo {author} {\bibfnamefont {C.~A.}\ \bibnamefont
  {Plata}}, \bibinfo {author} {\bibfnamefont {A.}~\bibnamefont {Prados}},\ and\
  \bibinfo {author} {\bibfnamefont {E.}~\bibnamefont {Trizac}},\ }\bibfield
  {title} {\bibinfo {title} {Driving rapidly while remaining in control:
  classical shortcuts from {Hamiltonian} to stochastic dynamics},\ }\href@noop
  {} {\bibfield  {journal} {\bibinfo  {journal} {Rep. Prog. Phys.}\ } (\bibinfo
  {year} {2022})}\BibitemShut {NoStop}%
\bibitem [{\citenamefont {Polettini}\ and\ \citenamefont
  {Esposito}(2013)}]{Polettini}%
  \BibitemOpen
  \bibfield  {author} {\bibinfo {author} {\bibfnamefont {M.}~\bibnamefont
  {Polettini}}\ and\ \bibinfo {author} {\bibfnamefont {M.}~\bibnamefont
  {Esposito}},\ }\bibfield  {title} {\bibinfo {title} {Nonconvexity of the
  relative entropy for {Markov} dynamics: A {Fisher} information approach},\
  }\href {https://doi.org/10.1103/PhysRevE.88.012112} {\bibfield  {journal}
  {\bibinfo  {journal} {Phys. Rev. E}\ }\textbf {\bibinfo {volume} {88}},\
  \bibinfo {pages} {012112} (\bibinfo {year} {2013})}\BibitemShut {NoStop}%
\bibitem [{\citenamefont {Maes}\ \emph {et~al.}(2011)\citenamefont {Maes},
  \citenamefont {Neto\ifmmode~\check{c}\else \v{c}\fi{}n\'y},\ and\
  \citenamefont {Wynants}}]{Maes}%
  \BibitemOpen
  \bibfield  {author} {\bibinfo {author} {\bibfnamefont {C.}~\bibnamefont
  {Maes}}, \bibinfo {author} {\bibfnamefont {K.}~\bibnamefont
  {Neto\ifmmode~\check{c}\else \v{c}\fi{}n\'y}},\ and\ \bibinfo {author}
  {\bibfnamefont {B.}~\bibnamefont {Wynants}},\ }\bibfield  {title} {\bibinfo
  {title} {Monotonic return to steady nonequilibrium},\ }\href
  {https://doi.org/10.1103/PhysRevLett.107.010601} {\bibfield  {journal}
  {\bibinfo  {journal} {Phys. Rev. Lett.}\ }\textbf {\bibinfo {volume} {107}},\
  \bibinfo {pages} {010601} (\bibinfo {year} {2011})}\BibitemShut {NoStop}%
\bibitem [{\citenamefont {Gladrow}\ \emph {et~al.}(2019)\citenamefont
  {Gladrow}, \citenamefont {Ribezzi-Crivellari}, \citenamefont {Ritort},\ and\
  \citenamefont {Keyser}}]{Felix_2019}%
  \BibitemOpen
  \bibfield  {author} {\bibinfo {author} {\bibfnamefont {J.}~\bibnamefont
  {Gladrow}}, \bibinfo {author} {\bibfnamefont {M.}~\bibnamefont
  {Ribezzi-Crivellari}}, \bibinfo {author} {\bibfnamefont {F.}~\bibnamefont
  {Ritort}},\ and\ \bibinfo {author} {\bibfnamefont {U.~F.}\ \bibnamefont
  {Keyser}},\ }\bibfield  {title} {\bibinfo {title} {Experimental evidence of
  symmetry breaking of transition-path times},\ }\bibfield  {journal} {\bibinfo
   {journal} {Nat. Commun.}\ }\textbf {\bibinfo {volume} {10}},\ \href
  {https://doi.org/10.1038/s41467-018-07873-9} {10.1038/s41467-018-07873-9}
  (\bibinfo {year} {2019})\BibitemShut {NoStop}%
\end{thebibliography}%

\end{document}